# Nonlinear optical spectrum of diamond at femtosecond regime

Juliana M. P. Almeida*, Charlie Oncebay, Jonathas P. Siqueira, Sérgio R. Muniz, Leonardo De Boni, Cleber R. Mendonça**

São Carlos Institute of Physics, University of São Paulo, PO Box 369, 13560-970, São Carlos, SP, Brazil.

* julianamara@ifsc.usp.br **crmendon@ifs.usp.br

**ABSTRACT**

Although diamond photonics has driven considerable interest and useful applications, as shown in frequency generation devices and single photon emitters, fundamental studies on the third-order optical nonlinearities of diamond are still scarce, stalling the development of an integrated platform for nonlinear and quantum optics. The purpose of this paper is to contribute to those studies by measuring the spectra of two-photon absorption coefficient (β) and the nonlinear index of refraction ($n_2$) of diamond using femtosecond laser pulses, in a wide spectral range. These measurements show the magnitude of β increasing from 0.07 to 0.23 cm/GW, as it approaches the bandgap energy, in the region from 3.18 to 4.77 eV (390 – 260 nm), whereas the $n_2$ varies from zero to $1.7 \times 10^{-19}$ m²/W in the full measured range, from 0.83 - 4.77 eV (1500 – 260 nm). The experimental results are compared with theoretical models for nonlinear absorption and refraction in indirect gap semiconductors, indicating the two-photon absorption as the dominant effect in the dispersion of the third-order nonlinear susceptibility. These data, together with optical Kerr gate measurements, also provided here, are of foremost relevance to the understanding of ultrafast optical processes in diamond and its nonlinear properties.

# Introduction

Diamond presents a striking combination of mechanical, thermal and optical properties, which has prompted its investigation for the development of a new generation of photonic devices that are expected to overcome the silicon limitations [1]. Among its features are the wide transmission window, a low absorption loss and a high index of refraction, associated with excellent hardness and thermal conduction, all supporting the use of diamond in the design of integrated nonlinear and quantum optical platforms. Particularly, its ability to hold stable color centers, able of acting as single photon sources, has motivated many applications in quantum information science [1-4].

Recent advances on the fabrication of synthetic diamonds, along with good control of color centers have resulted in a significant progress towards the development of diamond Raman lasers [5,6] and quantum optics [1,4,7,8]. On the other hand, studies on the nonlinear optical properties of diamond are still limited, hindering the full development of integrated diamond photonics platforms [9]. Currently, the available data are restricted to a few measurements of optical nonlinearities at specific wavelengths [10-14], impairing the analysis of the spectral behavior of the optical nonlinearities.

Therefore, considering the interests on diamond nonlinear photonics, this paper presents measurements of third-order optical nonlinearities in type IIa diamond, performed over a broad spectral range, from 0.83 - 4.77 eV (1500 – 260 nm), at femtosecond regime. To the best of our knowledge, this is the first report of such a wide range of nonlinear optical properties in diamond, particularly for nonlinear refraction and absorption. The measurements were obtained by Z-scan technique are compared to theoretical prediction for an indirect gap semiconductors, which is based on phonon-assisted two-photon absorption[15]. In addition to the nonlinear refraction and absorption data, it is also presented here the optical Kerr gate data, in order to confirm the response time of the nonlinear refraction in diamond. Together, this measurements in such a wide spectral range are of foremost relevance to determine the correct order of magnitude of the two-photon absorption coefficient, still in debate in the literature, and unravel the dominant underlying mechanism responsible for the nonlinear refraction in diamond.

## Results and Discussions

Because diamond crystal lattice presents inversion symmetry, due to its cubic structure, second-order optical nonlinearities are absent and, therefore, third-order processes are the lowest order ones, being the most relevant for diamond photonics. Moreover, the recent interest on integrated diamond platforms for quantum and nonlinear optics [2,9] makes crucial the characterization of the third-order susceptibility of diamond in a wide wavelength range.

In this direction, we performed Z-scan measurements for diamond in the range of 0.83 - 4.77 eV (1500 - 260 nm). Typical Z-scan transmittance curves are shown in Fig. 1, for an excitation energy of 3.54 eV (350 nm) (first row) and 2.21 eV (560 nm) (second row). Figures 1(a) and 1(b) display, respectively, open and closed-aperture Z-scan data, along with the dependence of the transmittance change ($\Delta T$) versus the intensity (inset). The decrease in normalized transmittance observed in the nonlinear absorption curve (open aperture), shown in Fig. 1(a), indicates a two-photon absorption (2PA) process. The solid line in Fig. 1(a) represents the fit obtained using the standard procedure [16], from which we can determine the two-photon absorption coefficient β. Two-photon absorption was only observed for excitation energy ≥ 3.18 eV (λ ≤ 390 nm). For photon energies in which 2PA is significant, the refractive (closed aperture) Z-scan signature is asymmetric, with a valley greater than the peak, as displayed in Fig. 1(b). For excitation energies lower than 3.18 eV (λ > 390 nm), the refractive Z-scan curves become symmetrical, as illustrated in Fig. 1(d) representing an excitation at 2.21 eV (560 nm), indicating that the two-photon absorption process is not present anymore. This is confirmed by the lack of absorptive signature in Fig. 1(c). It is interesting to point out that the valley-peak configuration exhibited by the nonlinear refraction curves (Fig. 1(b) and 1(d)) are characteristic of positive $n_2$ values. The solid lines in Fig. 1 are the best fit, obtained according to Ref. [16], from which the nonlinear refractive index and nonlinear absorption coefficient can be obtained. The insets in Figures 1(a) and 1(b) show the linear behavior of the transmittance change (ΔT) as a function of the light intensity at 3.54 eV (350 nm), as expected for third-order nonlinearities.

From Z-scan curves similar to the ones displayed in Fig. 1, obtained at different excitation energy, the dispersion of β and $n_2$ can be obtained for the full spectral range. Figure 2 (closed circles) displays the experimental 2PA coefficient (β) spectrum of diamond. It is worth noting that 2PA was detected from 3.18 - 4.77 eV (390 - 260 nm), region where the photon energy ℏω is smaller than the band gap ($E_g$ = 5.54 eV) and higher than $E_g/2$. Notice that values of β increase from 0.07 to 0.23 cm/GW as the excitation energy approaches the bandgap energy.

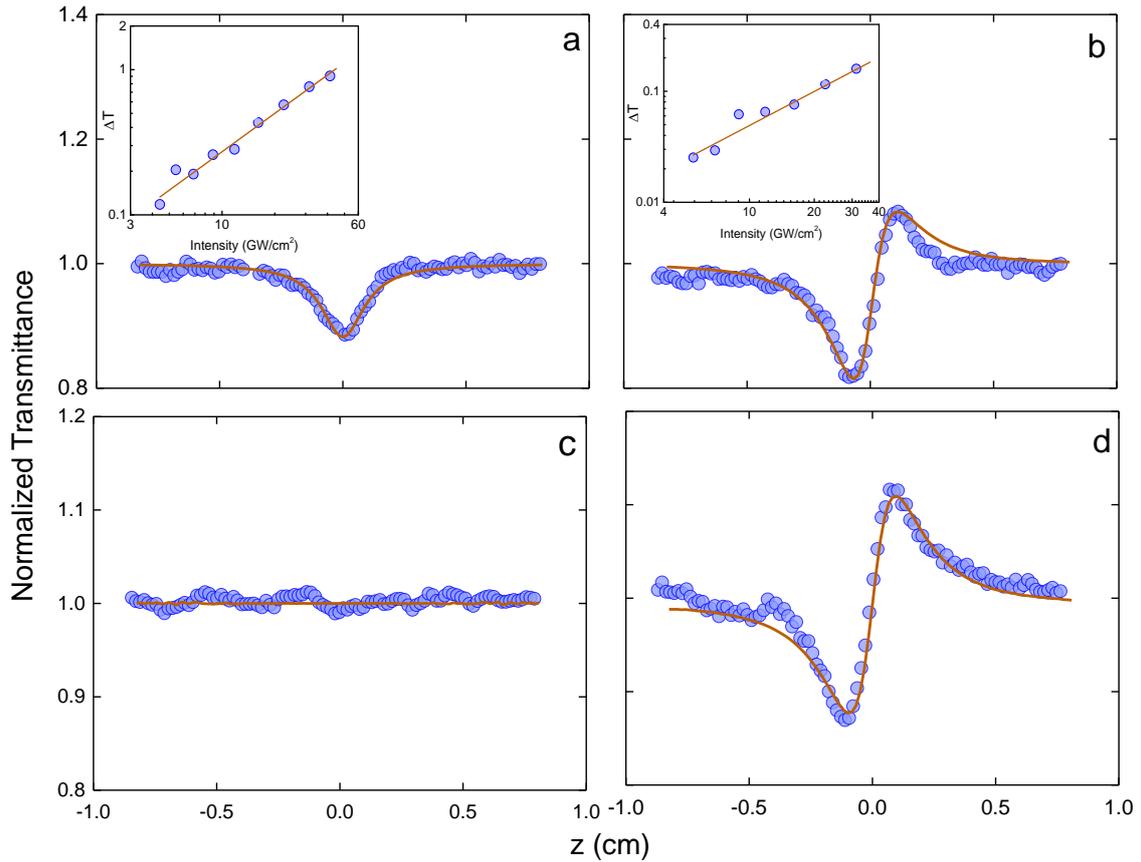

**Figure 1** – Open and closed Z-scan signatures at 3.54 eV (350 nm) (a and b, respectively) and 2.21 eV (560 nm) (c and d, respectively). The inset in (a) and (b) displays the linear dependence on the transmittance change (ΔT) with the intensity at 3.54 eV (350 nm), whose slopes are approximately 1 in the log-log plot.

A theory based on second-order perturbation developed by Sheik-Bahae et al. [17,18] has been successfully applied to modeling the dispersion of optical nonlinearities in wide-gap dielectrics and direct gap semiconductors. This theory does not include phonon-assisted transitions, a few models have been proposed for indirect gap materials [15,19,20], which is the case of diamond and other group IV semiconductors (Si and Ge). In particular, the formalism proposed by Garcia et al. [15] enables to predict the dispersion of the 2PA coefficient for indirect gap semiconductors, considering the sum of three transition processes: allowed-allowed (i = 0), allowed-forbidden (i = 1), and forbidden-forbidden (i = 2), according to $\beta(\omega) = \sum_{i=0}^{2} \beta^{(i)}(\omega)$, using

$$\beta^{(i)}(\omega) = K_i \frac{1}{n_0^2 E_{ig}^3} F_2^{(i)}\left(\frac{\hbar\omega}{E_{ig}}\right) \quad (1)$$

where $F_2^{(i)}(x) = \frac{(2x-1)^{i+2}}{(2x)^5}$ and $K_i$ is a curve-fitting factor, as described in Refs. [15,21], assuming the degenerate case. The linear refractive index ($n_0$) and the indirect band gap energy ($E_{ig}$) for diamond are, respectively, 2.4 and 5.54 eV. The solid line in Fig. 2 represents the fit obtained with Eq. (1), using $K_0$ = 62.86 × 10$^{-9}$, $K_1$ = 5.71 × 10$^{-9}$ m.eV$^3$/W and $K_2$ fixed as zero, once it has been shown that forbidden-forbidden transition can be neglected [15]. As it can be seen, there is a good agreement between experimental data and model, indicating that 2PA is, in fact, the major contribution to the observed nonlinear absorption.

The model given by Eq. (1) can be applied as a universal expression for indirect semiconductors, that enables predicting the 2PA coefficient in a wide spectral range. Therefore, it is interesting to compare the experimental results presented in Fig. 2 with the ones reported in literature, which are also plotted in Fig. 2 (open symbols and crosses) for better visualization [22]. As it can be seen, the previously reported data for 2PA coefficients, obtained using different techniques, varies significantly, from values in the same order of magnitude up to one order higher than the values determined herein (0.07 to 0.23 cm/GW) [10,23,24]. Such wide difference may arise from a combination of the temporal regime of laser pulses (eventually leading to contributions from thermal effects) and the number of free carriers in diamond. For instance, overestimated values of β may result from absorption processes not related to the third-order susceptibility, such as stepwise two-photon absorption processes, as in a cascade absorption by long-lived free carriers. This could prevent an accurate determination of β, if nor considered.

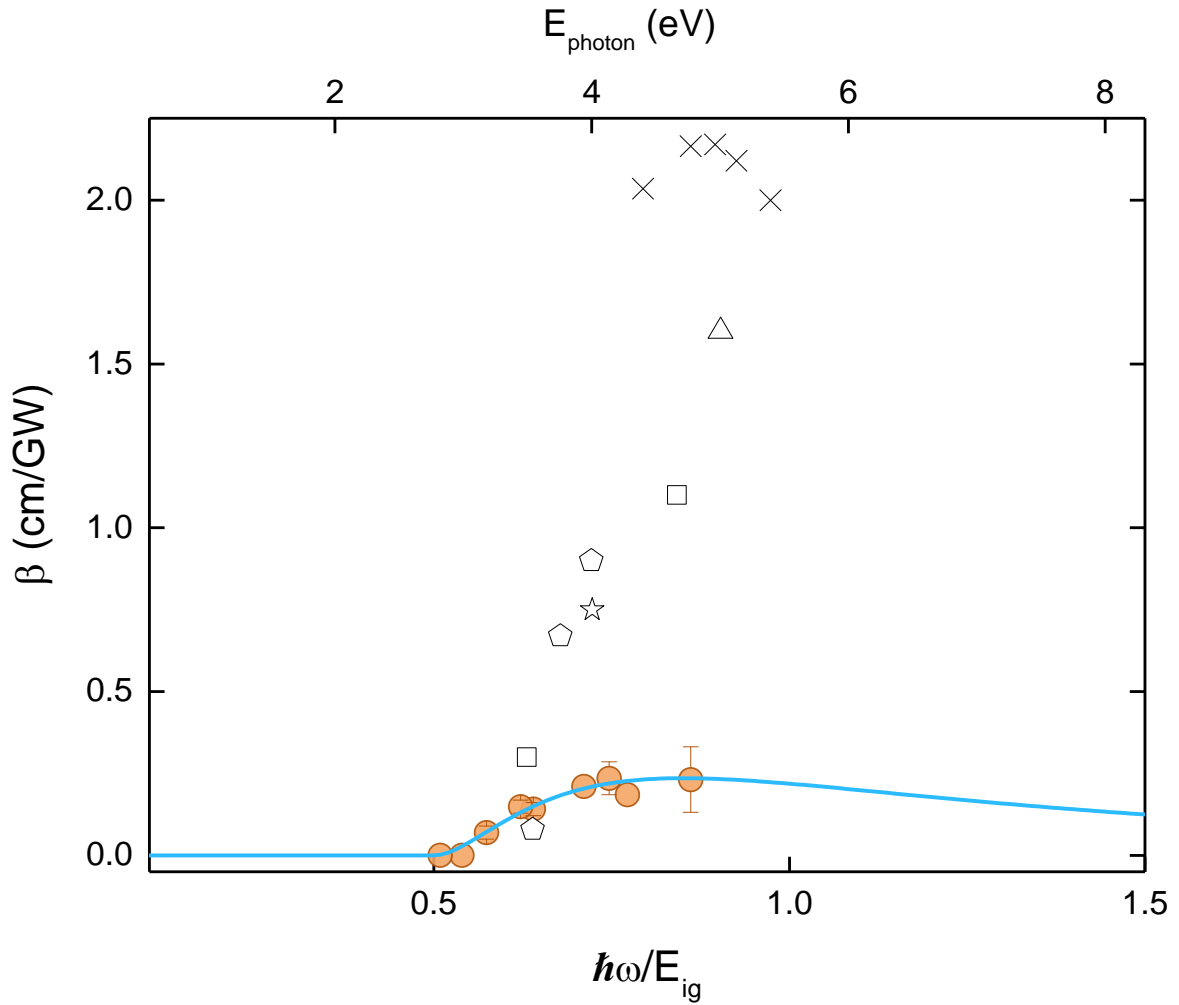

**Figure 2** - Dispersion of the two-photon absorption coefficient (β) of diamond. Solid circles correspond to the experimental data obtained in this work and continuous line is the theoretical fit obtained from Eq. 1. Literature data are plotted as crosses [24], up-triangles [26], squares [10], pentagons [11] and star [27].

M. Sheik-Bahae et al. [10] (squares in Fig. 2) have reported two values of β; at 3.49 eV (355 nm) they determined β= 0.3 cm/GW, which is similar to the one obtained here. However, at 4.66 eV (266 nm), much closer to the energy gap, the 2PA has increased drastically to a value of 1.1 cm/GW. According to their interpretation, such increase can be caused by re-absorption of long-lived free-carriers generated by the 2PA itself. Re-absorption by free-carriers is also enhanced by the long laser pulse widths, as it generates more population on the excited state. Gagarskii S. V. et al. [24] reported a value of one order of magnitude higher than ours (crosses in Fig. 2) and two times higher than the one reported in Ref. [10] for an excitation energy of 4.96 eV (250 nm). In ref. [24], the authors used pump-probe (100-fs pulses) and explained the high 2PA coefficient at the UV region as a cascade absorption of the probe (low-energy) due to the long-lived free carriers generated by the high-energy pump. As discussed in their paper, this cascade absorption acts as an additional process to the 2PA, thus increasing the overall nonlinear absorption, which leads to an overestimated 2PA coefficient in case the free-carrier absorption is not taken into account. Roth et al [23] measured the nonlinear absorption coefficient of diamond using femtosecond non-degenerated pump-probe, fixing the pump beam energy at 4.55 eV (273 nm), while tuning the probe beam energy between 0.22 and 3.00 eV (5500 and 410 nm); the 2PA coefficient was measured from 4.7 eV up to 7.5 eV, considering the sum of pump and probe beams. With such approach they were able to directly compare results obtained at different probe wavelengths. Also, they modeled the time-dependent transmission using rate equations, being able to discriminate contributions from 2PA and long lived free carriers' absorption, thus avoiding overestimating the 2PA coefficient. It is noteworthy that their results are in agreement with the ones presented herein, being lower than 0.5 cm/GW for most of the evaluated photon energies. The results of Roth et al [23] are not summarized in Fig. 2 because they deal with the non-degenerated case.

It is important to highlight that according to Eq. 1, 2PA coefficient scales with the inverse of the third-power of the band gap energy ($E_{ig}^{-3}$), therefore for materials with large band gap energy, as diamond, it is expected that 2PA be low when compared to other indirect semiconductors, as Si and Ge, for instance [21]. Thus, the large values reported on the literature for 2PA in diamond might be related to other nonlinear optical processes rather than pure two-photon absorption.

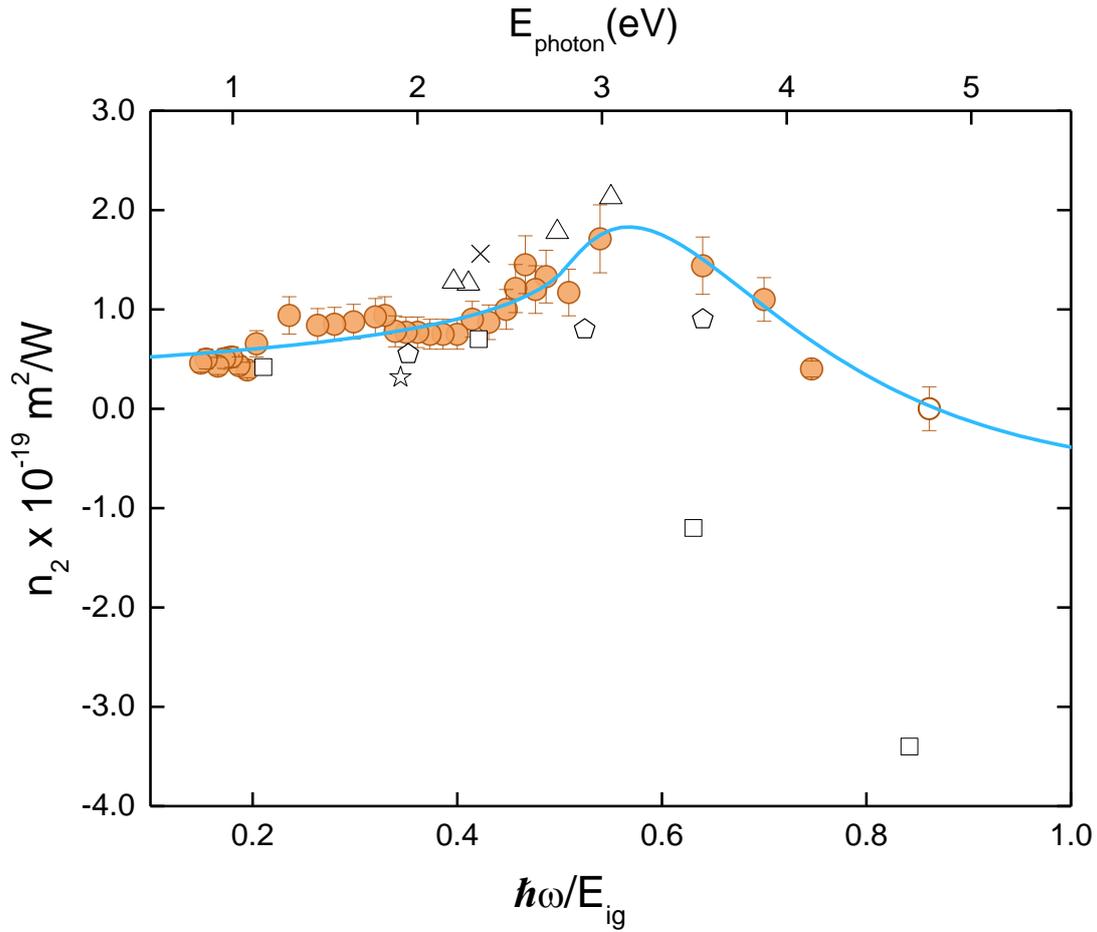

**Figure 3 -** Experimental (solid circles) and theoretical (solid line - Eq. 2) dispersion of the diamond nonlinear refractive index. Squares [10], pentagons [11], star [12], cross [13] and triangles [14] correspond to the values find in the cited references. The values in [14] were increased by a factor of 4 to follow the calculation and definition of $n_2$ used herein.

The nonlinear refractive index dispersion for diamond is shown in Fig. 3 (closed circles). The values of $n_2$, obtained from closed-aperture Z-scan measurements, vary from $0.4 \times 10^{-19}$ to $1.7 \times 10^{-19}$ m²/W, for the energy range of 0.83 - 4.13 eV ($\hbar\omega/E_{ig}$: 0.15 - 0.75; λ:1500 - 300 nm). The open circle in Fig. 3, at the right side of the graph, corresponds to the measurement performed at 4.77 eV ($\hbar\omega/E_g$ = 0.86; λ=260 nm), in which the Z-scan trace exhibits only nonlinear absorption feature; therefore the $n_2$ value was considered zero at this spectral position. Values of 1.3 × 10$^{-19}$ m$^2$/W and 8.2 × 10$^{-20}$ m$^2$/W have been reported for the visible and telecom regions, respectively [9]. Using Z-scan method and picosecond laser pulses, M. Sheik-Bahae et al. found the values illustrated in Fig. 3 by the open squares, in which $n_2$ > -34 × 10$^{-19}$ m$^2$/W at 4.66 eV (266 nm); -12 × 10$^{-19}$ m$^2$/W at 3.49 eV (355 nm); 7× 10$^{-19}$ m$^2$/W at 2.33 eV (532 nm) and 4.2 × 10$^{-19}$ m$^2$/W at 1.17 eV (1064 nm) [10], whereas $n_2$ ranging from (5.5 to 9) × 10$^{-20}$ m$^2$/W was determined within 1.95 to 3.54 eV for excitation with femtosecond laser pulses (pentagons in Fig. 3) [11]. Additional literature data are also displayed in Fig. 3 (open symbols), as summarized in ref. [22].

The Kerr effect in semiconductors can originate from absorptive processes presenting a quadratic dependence on the excitation field, such as 2PA, electronic Raman effect and optical Stark effect [17,18]. The Stark contribution is negative for all frequencies bellow the bandgap and becomes relevant only near the band edge, while the Raman contribution is small throughout the whole spectrum. Therefore, only the 2PA contribution was considered to model the nonlinear refraction evaluated herein, since it is the dominant effect on the dispersion of the third-order susceptibility ($\chi^3$). Thus, the solid line in Fig. 3 represents the modeling for the real part of $\chi^3$ based on the Kramers-Kronig relation and phonon-assisted two-photon absorption, using [15,21]

$$n_2(\omega) = \frac{c}{\pi} \int_0^\infty \frac{\beta(\omega,\omega')}{\omega'^2 - \omega^2} d\omega' \qquad (2)$$

whose constants were kept as described for Eq. 1 with exception to $K_0$ and $K_1$, which were increased by a factor of 3.5 for a better fitting. As it can be seen, there is a good agreement between the experimental data and the model represented in Eq. (2); with $n_2$ progressively increasing as $\hbar\omega/E_g$ is increased, reaching a maximum at approximately the 2PA absorption edge $\hbar\omega/E_g \sim 0.5$ (~3 eV; 415 nm), after which it decreases as the photon energy approaches the band gap. No negative values of $n_2$ were observed in our measurements, even for $\hbar\omega/E_g > 0.7$ (>3.88 eV; λ<320 nm), supporting the hypothesis that the most significant contribution to the nonlinear index of refraction dispersion comes from the 2PA term. It is worth noting, however, that in Ref. [10] it was observed a negative nonlinear refraction at 3.49 eV (355 nm) employing picosecond laser pulses, while here a clear positive $n_2$ is observed at 3.54 eV (350 nm), as shown in Fig. 1 (b). At the picosecond regime excited state contributions may affect the determination of a pure electronic Kerr response, as opposed to femtosecond laser pulses, thus explaining the discrepancy.

In order to confirm the response time of the nonlinear refraction of diamond optical Kerr gate (OKG) measurements were performed. Such measurements were carried out in different regions of the spectrum, specifically at 2.64, 2.48, 2.34, 2.18 and 1.97 eV (i.e. 470, 500, 530, 570 and 630 nm). Figure 4 displays only the OKG signal obtained at 2.34 eV (530 nm), because the same behavior was observed for the other wavelengths as well. This result reveals an instantaneous response time at the scale of the laser pulses used in our experiment (130 fs), revealing that the birefringence induced by the pump beam follows the pulse temporal shape. Therefore, the Kerr effect obtained herein for diamond corresponds to an ultrafast electronic optical nonlinearity.

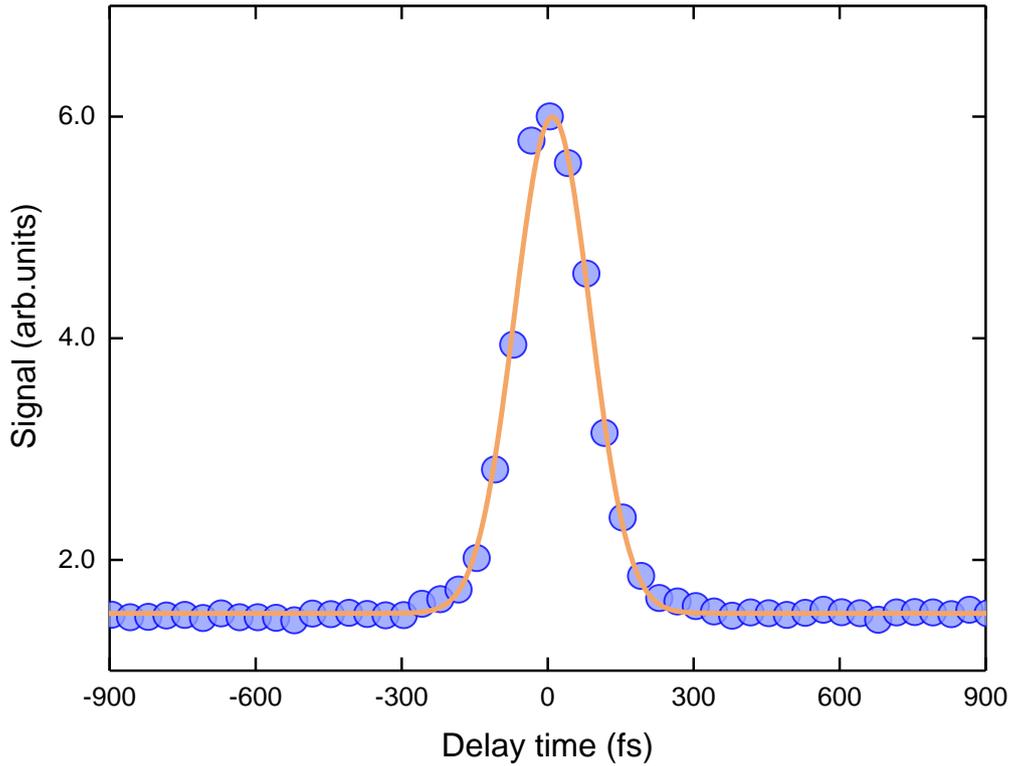

**Figure 4 -** Optical Kerr Gate signal obtained for the excitation of diamond at 2.34 eV (530 nm). The continuous line corresponds to a Gaussian fit, where one obtains the FWHM width of 130 fs that corresponds to the previously characterized pulse duration. Similar results were obtained for excitation at 2.64, 2.48, 2.18 and 1.97 eV (470, 500, 570 and 630 nm).

## Conclusions

Real and imaginary parts of the third-order nonlinear susceptibility of diamond, associated respectively with $n_2$ and $\beta$, have been investigated in the broad range 0.83 - 4.77 eV (1500 - 260 nm) using Z-scan measurements at femtosecond regime. Values of $\beta$ varies from 0.07 to 0.23 cm/GW, while $n_2$ is within zero and $1.7 \times 10^{-19}$ m²/W, both showing a good agreement with the theoretical modeling based on the phonon-assisted two-photon absorption process and Kramers-Kronig relations. The theoretical predictions reveal that the effect of two-photon absorption is dominant for the dispersion of $n_2$. Measurements of optical Kerr gate confirm that the nature of such optical nonlinearity arises from a pure electronic process, since it occurs in a time scale shorter than the pulse duration (130 fs). The determination of the third-order nonlinear spectrum of diamond reported herein is important for the development of an integrated nonlinear and quantum optics platform with applications in a wide spectral range.

## Methods

Third-order optical nonlinearities of a synthetic diamond, type-IIa single crystal sample ($2 \times 2 \times 0.53$ mm$^3$), with band gap energy of 5.54 eV ($\lambda_{cutoff}$ = 224 nm), was investigated using the Z-scan technique [16]. The commercially available sample was produced by CVD - chemical vapor deposition - and corresponds to Element Six's (E6) highest purity diamond, with a concentration of nitrogen impurities below 5 ppb (usually < 1 ppb) and concentration of nitrogen vacancies lower than 0.03 ppb. Confocal fluoresce imaging measurements of the sample showed no significant fluoresce, confirming the high purity. C Crystallography orientation is 100% single sector {100} and the polishing result in Ra<5nm on {100} face.

Open and closed aperture Z-scan measurements were performed to study the nonlinear absorption and the nonlinear refraction, respectively. Femtosecond laser pulses from a Ti:sapphire chirped pulse amplified system (150-fs, 775 nm and 1 kHz) were used to pump an optical parametric amplifier (OPA), which provides tunable 120-fs pulses from 0.62 to 2.7 eV (2000 - 460 nm). Depending on the wavelength region, $s$ or $p$ linear polarization was applied to the {100} crystal face.

For the Z-scan measurements, the beam from the OPA was spatially filtered and then focused in the sample, positioned in a translation stage, with a f = 15 cm quartz lens. A beam splitter was placed after the sample to provide a dual arm configuration, such that closed and open aperture Z-scan measurements were simultaneously carried out. The effect of the nonlinear absorption on the nonlinear refraction measurement was deducted by performing the ratio of the refractive (closed aperture) by the absorptive (open aperture) Z-scan signatures. Depending on the wavelength, the pulse energy and beam waist ranged, respectively, from 30 – 400 nJ and from 10 to 23 µm. Silicon or germanium photodetectors (according to excitation wavelength), coupled to a lock-in amplifier, were used to monitor the sample transmittance along its propagation, over a 16 mm scan-range. An additional setup of second-harmonic or third-harmonic generation was used to provide pulses from 2.7 eV up to 4.77 eV (460 nm down to 260 nm). The typical noise of the Z-scan signal is on the order of 1-5 %, which leads to a maximum experimental error of approximately 15% on the nonlinear coefficients determination. This estimate is based on comparison with fused silica measurements, which was used as reference material [25] to check the accuracy of the absolute values of $n_2$. The determination of the experimental uncertainty of β was performed using the standard deviation obtained for several measurements.

The time response of the nonlinearity was investigated using the optical Kerr gate (OKG), in which the same laser system and OPA were employed as the excitation source. Basically, in the OKG setup the beam is divided in a strong pump beam and a weak probe (4 % of the pump), which are focused and overlapped into the sample. The pump and probe beams are polarized at a 45° relative angle, in order to probe the birefringence induced by the pump beam due to a refractive index change on its polarization axis, as a result of the optical Kerr effect. The probe beam is time-delayed with respect to the pump, while its transmission through an analyzer is monitored, allowing to investigate the dynamics of the induced birefringence associated to the Kerr signal.

## Acknowledgments
The authors acknowledge the financial support from the Brazilian research funding agencies FAPESP (2015/17058-7, 2013/07276-1 and 2011/12399-0), CAPES and CNPq.

## Author Contributions Statement

J.M.P.A. performed the z-scan measurements at VIS-NIR and prepared the manuscript, J.P.S performed the z-scan measurements at UV region, C.O. and S.R.M conducted the theoretical analyses, L.D.B. accomplished the OKG experiment and C.R.M. leaded the discussion and analysis of all steps. All authors discussed the results and commented/revised the manuscript.

**Competing Interests:** The authors declare no competing financial interests.